\documentclass[preprint,number,sort&compress]{elsarticle}

\usepackage[utf8]{inputenc}
\usepackage[T1]{fontenc}
\usepackage[english]{babel}
\usepackage[unicode=true]{hyperref}
\usepackage{amsmath, amssymb, amsfonts}

\usepackage{graphicx}
\usepackage{float}
\usepackage{subcaption}
\usepackage{tikz}

\usepackage{bm}
\usepackage{enumerate}

\usepackage{hyperref}
\hypersetup{
    colorlinks=true,
    linkcolor=blue,
    citecolor=blue,
    urlcolor=blue,
    pdfstartview=FitV,
    breaklinks=true
}

\usepackage{orcidlink}
\usepackage{xcolor}

\begin{document}

\begin{frontmatter}

\title{Geometric Modeling of a Line of Alternating Disclinations: Application to Grain Boundaries in Graphene}

\author[ufal]{A. M. de M. Carvalho\,\orcidlink{0009-0006-3540-0364}}
\ead{alexandre@fis.ufal.br}

\author[ufpb]{C. Furtado\,\orcidlink{0000-0002-3455-4285}}
\ead{furtado@fisica.ufpb.br}

\address[ufal]{Instituto de Física, Universidade Federal de Alagoas, 57072-970, Maceió, AL, Brazil.}
\address[ufpb]{Departamento de Física, Universidade Federal da Paraíba, 58051-970, João Pessoa, PB, Brazil.}

\begin{abstract}
We develop a conformal geometric model for grain boundaries in graphene based on a periodic line of alternating disclinations. Within the framework of $(2+1)$-dimensional gravity, we solve a reduced form of the Einstein equations to determine the conformal factor, from which the induced metric, scalar curvature, and holonomy are obtained analytically. Each pentagon–heptagon pair is modeled as a disclination dipole, forming a continuous distribution that captures the geometric signature of experimentally observed 5|7 grain boundaries. We show that the curvature is localized near the defect line and that the geometry becomes asymptotically flat, with trivial holonomy at large distances. This construction provides a tractable and physically consistent realization of the Katanaev–Volovich framework, connecting topological defect theory with atomistic features of graphene.
\end{abstract}
\end{frontmatter}



\section{Introduction}

Topological defects play a fundamental role in the structure, dynamics, and emergent properties of condensed matter systems. Among them, disclinations — line defects associated with angular deficits or excesses in the local ordering — have long been recognized as elementary topological excitations in two-dimensional materials and anisotropic fluids~\cite{KlemanLavrentovich2003,Nelson2002}. These defects are not merely local imperfections: they encode geometric frustration, act as localized sources of curvature and holonomy~\cite{Furtado2008}, and can significantly alter the global topology of the underlying medium. In many systems, disclinations and other topological defects assemble into regular networks; notable examples include vortex lattices in superfluids and Bose–Einstein condensates, flux line arrays in type-II superconductors, and disclination structures in nematic and cholesteric phases~\cite{Berche2020,Fumeron2023b}. Their presence underpins a wide range of physical phenomena and opens the door to geometric control of material properties.
Originally formulated through the Volterra process~\cite{PuntigamSoleng1997} and later extended to incorporate plastic relaxation and continuous symmetries~\cite{Kleman2008}, disclinations have proven to be versatile structures mediating between geometry and material response. They arise not only in crystalline solids and liquid crystalline phases, but also in biological tissues, soft matter, liquid crystals and cosmological models~\cite{Fumeron2022,Fumeron2023b}.

In graphene and other two-dimensional materials, disclinations manifest as pentagonal or heptagonal rings and play a central role in the formation of grain boundaries. As highlighted in recent geometric formulations~\cite{Fumeron2023b}, the curvature sourced by these defects can be exploited to modulate electronic and transport properties, enabling novel strategies for the design of functional materials through defect engineering.

Graphene, a single layer of carbon atoms arranged in a honeycomb lattice, offers a paradigmatic platform for investigating topological defects. While often idealized as a perfectly regular hexagonal structure~\cite{Novoselov2004,Geim2007}, real samples frequently display structural irregularities, including pentagon–heptagon pairs that arise at the interfaces between misoriented crystalline domains—known as grain boundaries (GBs). These 5|7 defects correspond to positive and negative disclinations, respectively, and emerge naturally during growth processes or mechanical deformation~\cite{Hashimoto2004,Ma2014}. Atomistic simulations and high-resolution microscopy~\cite{YazyevLouie2010,LuicanMayer2016} have revealed that GBs in graphene are typically composed of periodic sequences of such disclination dipoles, giving rise to long-range strain fields and locally altering electronic behavior~\cite{CastroNeto2009}. Experimental STM imaging of GBs on SiC substrates has confirmed the presence of periodic 5|7 arrangements and associated curvature effects~\cite{Tison2014}, reinforcing the physical relevance of the geometric patterns modeled here. In many of these configurations, each 5|7 pair can be viewed as a dipole of curvature sources, and the entire GB behaves as a quasi-one-dimensional structure modulating the geometry of the graphene sheet. While numerical and empirical models have provided important insights into these systems, an explicit analytical treatment capable of describing the full metric and curvature field generated by periodic disclination arrays — and directly comparable with atomistic results — remains lacking.

The continuum geometric interpretation of these defects has been formalized in works such as those by Katanaev and Volovich~\cite{Katanaev1992}, where crystalline defects are described using curvature and torsion within the framework of Riemann–Cartan geometry. In this approach, disclinations appear as localized sources of curvature, while dislocations are associated with torsion, providing a unified geometric description of topological defects through differential geometry~\cite{Nakahara2003}.

While numerical and atomistic models have provided detailed insights into the energetics and structure of grain boundaries, an explicit analytical formulation capable of reproducing the geometry and topology of periodic disclination arrays — and directly comparable to atomistic data — remains largely undeveloped.

Beyond continuum and geometric approaches, atomistic simulations have been extensively used to study grain boundary structures and misorientation effects in graphene and related materials~\cite{riet2020}. In addition, large-scale atomistic methods have successfully captured defect dynamics in broader material systems, such as helium bubble growth in complex palladium alloys~\cite{wong2016helium}. These complementary perspectives underscore the value of integrating atomistic insights with analytical geometric frameworks.


In this work, we develop an analytical geometric model for grain boundaries in graphene, based on a periodic array of alternating disclinations. Adopting a conformal metric approach, we solve the Poisson equation for the conformal factor sourced by a dipole lattice and derive the resulting metric deformation, curvature distribution, and holonomy structure. We demonstrate that this model captures essential features of realistic GBs in graphene, including curvature localization, misorientation angles, and the energy scaling observed in atomistic simulations~\cite{Malola2010,Lorentzen2024}.

This geometric framework establishes a connection between discrete atomistic representations and smooth continuum descriptions, offering deeper insight into the interplay between topology, geometry, and material response. It enables a continuous interpolation across different GB configurations and provides a physically motivated analytic tool to investigate the effects of periodic defects. The approach also lays the groundwork for future studies on elastic backreaction, curvature-driven electronic phenomena, and defect-based material design in two-dimensional systems.

The goal of this paper is to construct a geometric description of grain boundaries that links curvature, holonomy, and misorientation in a unified Riemannian framework, thereby providing a natural bridge between continuum geometry and defect-based material design in two-dimensional systems.

\section{Geometric Theory of Topological Defects}
Crystalline defects in two-dimensional systems can be effectively described within a differential geometric framework, in which disclinations and dislocations are represented as localized sources of curvature and torsion, respectively, embedded in a Riemannian or Riemann–Cartan manifold~\cite{Katanaev1992,Nelson2002}. Rotational mismatches give rise to curvature (disclinations), while translational mismatches lead to torsion (dislocations). This formulation recasts elasticity theory as a covariant field theory, providing analytical tools to explore geometric effects beyond the reach of atomistic simulations.

The geometric theory of defects developed by Katanaev and Volovich~\cite{Katanaev1992} provides a unified framework for modeling topological defects in elastic media using Riemann--Cartan geometry. In this approach, the medium is treated as a three-dimensional manifold equipped with a metric \( g_{ij} \), torsion \( T^i{}_{jk} \), and curvature \( R^i{}_{jkl} \), which together encode both the elastic deformations and the topological structure of defects. Disclinations act as sources of curvature, while dislocations generate torsion.

A key advantage of this formalism is that it replaces singular boundary conditions and discontinuities with smooth geometric fields, allowing both localized and distributed defects to be described on equal footing. This geometrization is compatible with local symmetries and variational principles, and it reformulates complex physical problems as differential equations for the underlying geometry—such as Poisson equations for the conformal factor or three-dimensional Einstein-like equations with defect sources.

To model defects within this geometric framework, we express the two-dimensional metric in conformal form:
\begin{equation}
    ds^2 = e^{2\Omega(x,y)}(dx^2 + dy^2),
    \label{mconforme}
\end{equation}
where the conformal factor \(\Omega(x,y)\) encodes the geometric properties of the defect. Typically, \(\Omega(x,y)\) attains values of order unity near the disclination cores and decays rapidly away from the grain boundary, consistent with the localization of curvature observed in atomistic simulations. This behavior ensures that the geometry is strongly deformed in the vicinity of the defects, while recovering approximate flatness far from the boundary. The underlying geometry must satisfy an effective field equation analogous to Einstein’s equation in two dimensions:

\begin{equation}
    R_{\mu\nu} - \frac{1}{2}g_{\mu\nu}R = -8\pi G T_{\mu\nu},
\label{einstein}
\end{equation}
where \(R_{\mu\nu}\) is the Ricci tensor, \(R\) the scalar curvature, \(g_{\mu\nu}\) is the metric tensor, \(T_{\mu\nu}\) an effective energy-momentum tensor representing the distribution of defects, and \(G\) a coupling constant associated with the elastic properties of the medium. In this context, \(T_{\mu\nu}\) does not represent physical energy, but rather the stress and strain fields generated by topological defects, with the component \(T_{zz}\) encoding the density of angular deficits. This formal analogy is further explored in defect-based models formulated within the Katanaev–Volovich theoretical scheme~\cite{Katanaev1992},

The Ricci scalar associated with the conformal metric(\ref{mconforme}) can be written as
\begin{equation}
    R(x,y) = -e^{-2\Omega(x,y)} \Delta \Omega(x,y),
\end{equation}
where \(\Delta\) is the Laplacian in flat two-dimensional space. Substituting into the trace of the Einstein-like equation~\ref{einstein} leads to the following Poisson equation for the conformal factor:
\begin{equation}
    \Delta \Omega(x,y) = -\lambda(x,y).
    \label{poisson}
\end{equation}
Physically, \( \lambda(x,y) \) represents the spatial distribution of curvature induced by the defect array, with magnitude proportional to the strength of the disclination dipoles. In our model, it is chosen to reflect the periodicity along the grain boundary and the exponential localization transverse to it.

In this work, we adopt this conformal geometric approach to model grain boundaries in graphene. These boundaries, observed experimentally and reproduced in atomistic simulations~\cite{YazyevLouie2010,Malola2010,Lorentzen2024}, are composed of periodic sequences of 5|7 disclination dipoles aligned along a line. We model such boundaries by prescribing a periodic distribution of dipolar curvature sources \( \lambda(x,y) \), solve the resulting Poisson equation~(\ref{poisson})  and derive the induced metric and curvature. This formulation allows us to describe the local curvature peaks, asymptotic behavior, and holonomy induced by grain boundaries, providing a bridge between atomistic structure and continuous geometry.

\section{Solution of the Conformal Factor}

Grain boundaries in graphene often consist of periodic arrangements of non-hexagonal rings, typically alternating pentagons and heptagons. These structures can be interpreted geometrically as disclination dipoles: pairs of opposite curvature sources—positive for pentagons and negative for heptagons—separated by a finite distance~\cite{KlemanLavrentovich2003,Hashimoto2004}. Although they locally disrupt the sixfold symmetry of the lattice, they preserve atomic coordination and connectivity, making them ideal candidates for a continuum description based on curvature. In this picture, the vector connecting the disclinations defines the Burgers vector of the associated dislocation, establishing a direct link between lattice translations and geometric charge.

We model a grain boundary as a periodic array of such disclination dipoles of strength \( +m \) and \( -m \), aligned along the \( x \)-axis. The parameter \( m \) quantifies the angular strength of each disclination and is proportional to the Frank angle associated with the topological defect. In graphene, a typical value is \( m = \pm \frac{\pi}{3} \), corresponding to the ±60° rotation observed in adjacent pentagon-heptagon pairs. This configuration captures the essential features observed in first-principles simulations~\cite{Hashimoto2004,YazyevLouie2010,Malola2010,Lorentzen2024}, where extended 5|7 defect lines appear as stable grain boundary motifs. In the continuum limit, this leads to a periodically modulated distribution of curvature sources confined to a defect line, suitable for analytical treatment within the conformal geometry framework.

To model a periodic array of disclination dipoles, we prescribe the curvature source as
\begin{eqnarray}
    \lambda(x,y) &=& \sum_{n=-\infty}^{\infty} \left[ m \, \delta(x - na - a/2)\delta(y) - m \, \delta(x - na + a/2)\delta(y) \right],
\end{eqnarray}

In this context, the parameter \( a \) represents the periodic separation between adjacent disclination dipoles. In realistic graphene grain boundaries, \( a \) typically ranges from \( 1\,\text{nm} \) to \( 2\,\text{nm} \), consistent with the inter-disclination spacing observed in atomistic models and experimental microscopy studies~\cite{YazyevLouie2010,Malola2010}.


This configuration reflects the translational symmetry observed in grain boundaries and ensures topological neutrality. The parameter \( m \) represents the strength of each disclination and is directly related to the geometric deficit or excess angle. Specifically, \( m = \pm 1 \) corresponds to elementary defects with angular mismatch of \( \pm \frac{\pi}{3} \), which naturally arise in graphene as isolated pentagons or heptagons within a hexagonal lattice.
To solve the Poisson equation with this source~\ref{poisson}, we apply the method of Green's functions. The solution for \(\Omega(x,y)\) is given by the convolution~\cite{CourantHilbert1962,Jackson1999,MorseFeshbach1953,Arfken2013}
\begin{eqnarray}
    \Omega(x,y) &=& \int G(x - x', y - y') \, \lambda(x', y') \, dx' dy',
\end{eqnarray}
where \(G(x,y)\) is the Green's function of the two-dimensional Laplacian with periodic boundary conditions along \(x\). The appropriate expression is
\begin{eqnarray}
    G(x,y) &=& -\frac{1}{4\pi} \ln \left[ \cosh\left( \frac{2\pi y}{a} \right) - \cos\left( \frac{2\pi x}{a} \right) \right],
\end{eqnarray}
which satisfies
\begin{eqnarray}
    \Delta G(x,y) &=& \sum_{n=-\infty}^{\infty} \delta(x - na)\delta(y) - \frac{1}{a} \delta(y).
\end{eqnarray}

Since our source \(\lambda(x,y)\) is a dipolar combination of such periodic delta functions, it automatically satisfies the neutrality condition, and the second term in the Green function cancels out upon convolution. The resulting conformal factor becomes
\begin{eqnarray}
    \Omega(x,y) &=& -\frac{m}{4\pi} \ln \left[ \cosh\left( \frac{2\pi y}{a} \right) - \cos\left( \frac{2\pi x}{a} \right) \right],
\label{fconforme}
\end{eqnarray}
with the prefactor \(-\frac{m}{4\pi}\) ensuring the correct normalization, consistent with the Green’s function for a two-dimensional Poisson equation.

This conformal factor also encodes both the periodicity of the disclination array and the exponential decay of the metric deformation away from the grain boundary. Its logarithmic structure ensures regular behavior away from the defect cores, while the symmetry of the solution reflects the alternating nature of the 5|7 pairs. Thus captures the essential geometric features observed in atomistic models~\cite{YazyevLouie2010,Lorentzen2024}, including curvature localization, misorientation, and periodic strain modulation along the boundary. These features are consistent with the polygonal 5|7 motifs and periodic strain textures observed in atomistic studies~\cite{Malola2010}.

Overall, the conformal factor describes a smooth, localized geometric deformation with exponential decay, in agreement with the spatial confinement of strain and curvature observed in both theoretical models and atomistic simulations of graphene grain boundaries~\cite{Malola2010,Lorentzen2024}. This asymptotic profile will be essential for interpreting curvature localization and topological neutrality in the context of the continuum theory.

\begin{figure}[htbp]
    \centering
    \includegraphics[width=0.75\textwidth]{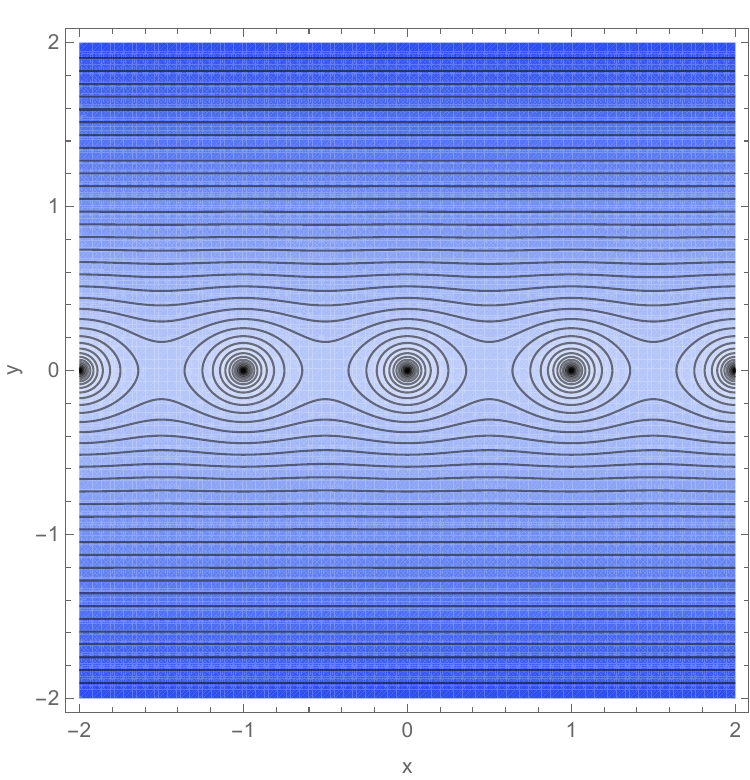}
    \caption{Contour map of the conformal factor \(\Omega(x,y)\) associated with a periodic line of alternating disclinations. The dashed red line indicates the grain boundary (\(y=0\)), while black dots mark the disclination cores with alternating signs \(+m\) and \(-m\).}
    \label{fig:omega_contour_grain}
\end{figure}

\begin{figure}[htbp]
    \centering
    \includegraphics[width=0.75\textwidth]{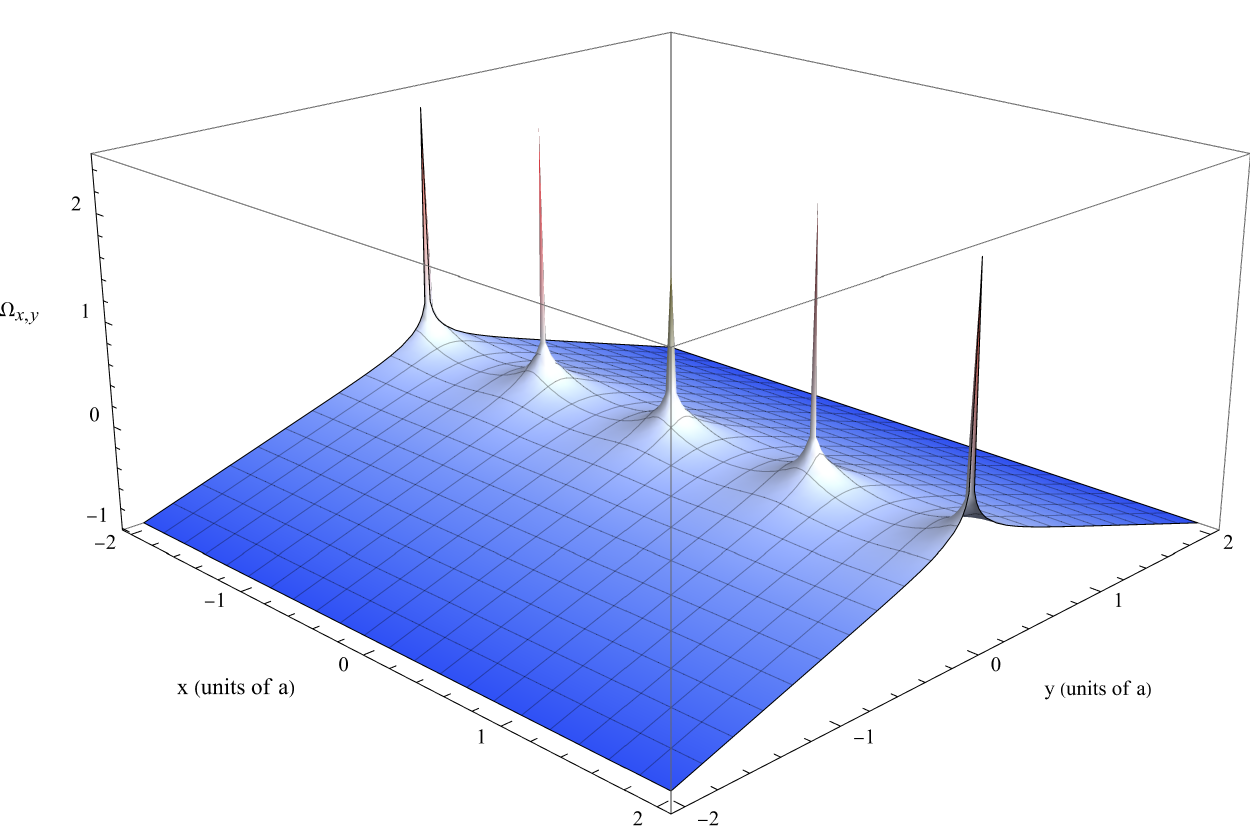}
    \caption{3D surface plot of the conformal factor \(\Omega(x,y)\), highlighting the geometric valleys and peaks induced by the periodic disclination array. The transverse direction \(y\) shows strong modulation with exponential decay away from the grain boundary.}
    \label{fig:omega_3d_grain}
\end{figure}

Figures~\ref{fig:omega_contour_grain} and~\ref{fig:omega_3d_grain} illustrate the geometric structure encoded by the conformal factor \(\Omega(x,y)\) for a periodic grain boundary modeled as a sequence of alternating disclinations. The contour map reveals the localized deformation near the disclination cores, while the 3D surface plot emphasizes the intensity and confinement of geometric deformation along the transverse direction. These visualizations confirm the spatial periodicity and anisotropic character of the conformal metric, supporting its role as an effective continuum description of disclinated grain boundaries in 2D materials.


\section{Asymptotic Behavior and Far-field Limit}

In this section, we analyze the asymptotic behavior of the conformal factor \(\Omega(x,y)\), focusing on its decay far from the grain boundary. This allows us to understand how the geometric deformation induced by the periodic array of disclination dipoles localizes in space. We recall that the conformal factor corresponding to a periodic array of disclination dipoles is given by a logarithmic expression involving hyperbolic and trigonometric functions~\cite{YazyevLouie2010,Lorentzen2024}. In the far-field regime, where \( |y| \gg a \), we apply the asymptotic expansion

\begin{eqnarray}
    \cosh\left( \frac{2\pi y}{a} \right) &\sim& \frac{1}{2} e^{2\pi |y|/a},
\end{eqnarray}
to simplify the argument of the logarithm:
\begin{eqnarray}
    \Omega(x,y) &\approx& -\frac{m}{4\pi} \ln\left[ \frac{1}{2} e^{2\pi |y|/a} \left( 1 - 2 \cos\left( \frac{2\pi x}{a} \right) e^{-2\pi |y|/a} + \mathcal{O}(e^{-4\pi |y|/a}) \right) \right] \nonumber \\
    &=& -\frac{m}{4\pi} \left[ \frac{2\pi |y|}{a} + \ln\left( \frac{1}{2} \right) - 2 \cos\left( \frac{2\pi x}{a} \right) e^{-2\pi |y|/a} + \cdots \right].
\end{eqnarray}

Thus, the leading-order behavior of the conformal factor becomes~\ref{fconforme}
\begin{eqnarray}
    \Omega(x,y) &\sim& -\frac{m}{2a} |y| + \text{const} + \mathcal{O}(e^{-2\pi |y|/a}),
\end{eqnarray}
showing that \(\Omega(x,y)\) decays linearly with \(|y|\) in the far field. The associated conformal metric takes the asymptotic form
\begin{eqnarray}
    ds^2 &\to& e^{-m |y|/a} (dx^2 + dy^2), \quad \text{as } |y| \to \infty,
\end{eqnarray}
The resulting metric resembles that of the Lobachevsky (hyperbolic) plane in quasi-Cartesian coordinates~\cite{doCarmo1976,ONeill1983}, featuring an exponential prefactor that signals negative curvature in the transverse direction. Unlike the hyperbolic plane, however, the curvature in our case decays rapidly and integrates to zero over the full domain, ensuring asymptotic flatness~\cite{Katanaev1992,Gibbons2008}. A relevant point is that the notion of “geometric confinement” is appropriate here, but it should be noted that this does not correspond to a constant asymptotic curvature (as in the hyperbolic plane), but rather to a localized band of curvature confined around \( y \sim 0 \).

This asymptotic behavior supports the interpretation of the grain boundary as a confined geometric defect — a localized strip of curvature embedded in an otherwise flat space. The dominant contribution to \(\Omega(x,y)\) is independent of \(x\), implying that the spatial periodicity becomes negligible at large \(|y|\). The subleading correction, captures the periodic modulation induced by the disclination dipole array and becomes relevant only near the defect line. This modulation plays a key role in the local structure of the curvature field, as will be analyzed in the following section.

\section{Curvature Profile and Topological Structure}

The conformal factor derived in the previous section determines the local geometry of the space through the scalar curvature. This section presents the explicit expression for the curvature, analyzes its spatial structure, and explores its topological implications. The results reveal how a periodic arrangement of disclination dipoles generates a rich and localized curvature profile, consistent with atomistic observations in graphene grain boundaries~\cite{YazyevLouie2010,Malola2010,Lorentzen2024}. Although the disclinations are discrete, the resulting scalar curvature is continuous and exhibits a periodic structure concentrated near the grain boundary, where the geometric deformation is most significant.

Using the explicit form of the conformal factor~\ref{fconforme}, we obtain the following expression for the curvature:
\begin{eqnarray}
R(x, y) &=& \frac{8\pi^2 m}{a^2} \left[ \frac{\cos\left(\frac{2\pi x}{a}\right)\cosh\left(\frac{2\pi y}{a}\right) - 1}{\left(\cosh\left(\frac{2\pi y}{a}\right) - \cos\left(\frac{2\pi x}{a}\right)\right)^2} \right] \nonumber \\
&+& \frac{4\pi^2 m^2}{a^2} \left[ \frac{\sin^2\left(\frac{2\pi x}{a}\right)}{\left(\cosh\left(\frac{2\pi y}{a}\right) - \cos\left(\frac{2\pi x}{a}\right)\right)^2} \right].
\label{RicciR}
\end{eqnarray}

This expression contains two contributions: a term linear in $m$ arising from
the Laplacian of $\Omega$, and a nonlinear term proportional to $m^2$ due to the
conformal prefactor $e^{-2\Omega}$. This highlights the interplay between
curvature and the strength of the disclination dipoles, reflecting how geometric
effects become more pronounced for larger $m$.

The scalar curvature \( R(x,y) \) measures the local deviation from flatness and peaks sharply at the grain boundary core, reflecting the geometric distortion induced by the periodic disclination array.

\begin{figure}[H]
    \centering
    \includegraphics[width=0.5\textwidth]{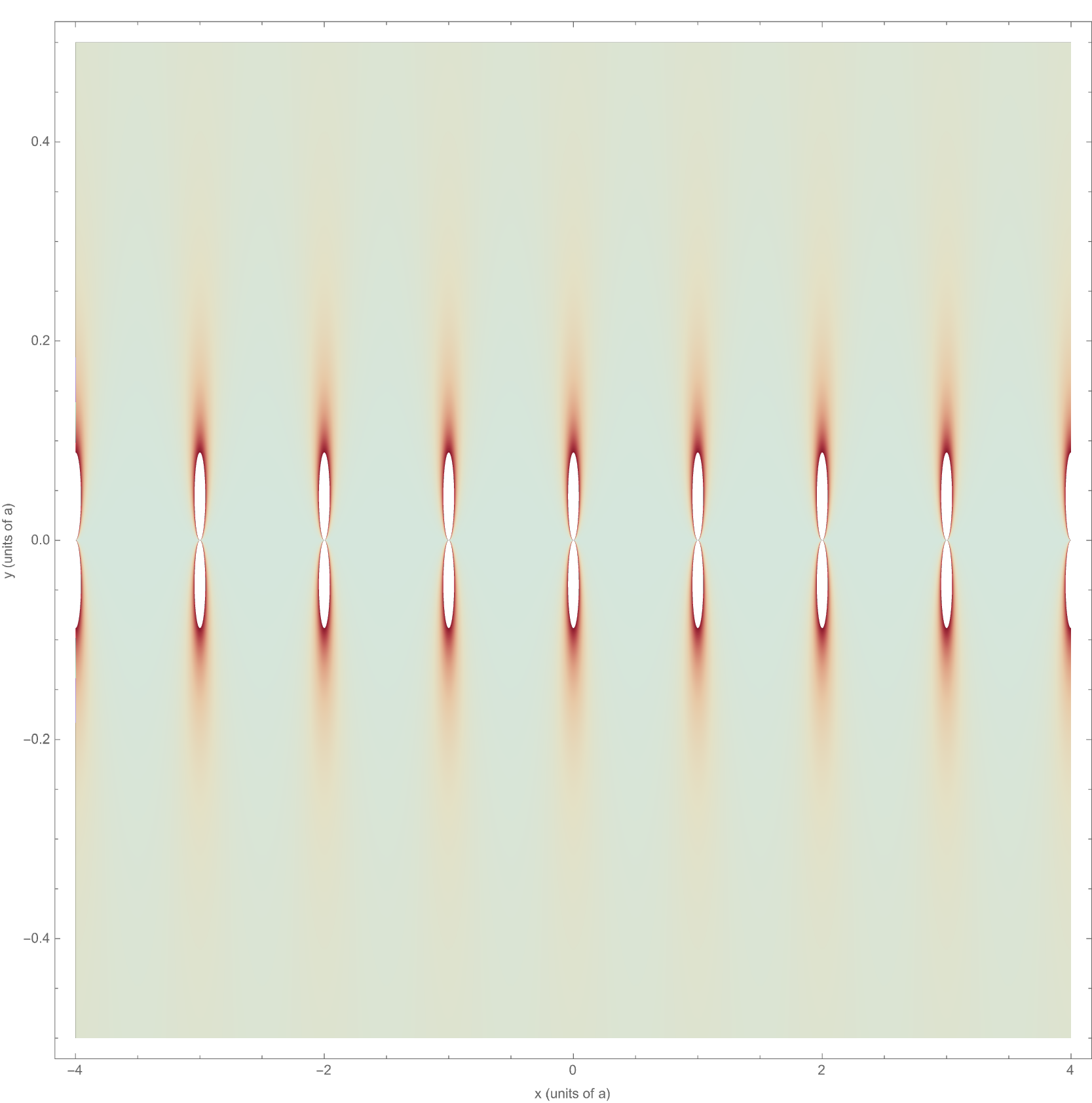}
   \caption{Heatmap of the scalar curvature $R(x, y)$ generated by a periodic line of disclination dipoles (5--7 pairs) in graphene. The plot highlights the symmetric, periodic structure and the confinement of curvature to a narrow region around the grain boundary.}

    \label{fig:curvatureHeatmap}
\end{figure}
Figure~\ref{fig:curvatureHeatmap} shows the heatmap of the scalar curvature $R(x, y)$ for a periodic array of disclination dipoles in graphene, modeled according to the conformal geometric framework. The curvature is not sharply localized at single points, but rather distributed in symmetric elliptical regions along the defect line at $y = 0$. This shape reflects the smooth geometric distortion generated by 5--7 disclination pairs. The curvature profile displays exponential decay in the transverse ($y$) direction and periodic modulation along the $x$-axis, in full agreement with the analytical expression of $R(x, y)$. The elliptical shape of the high-curvature regions confirms the geometric confinement induced by the defect array and supports the interpretation of the grain boundary as a localized and topologically neutral curvature texture in the continuum limit.


The scalar curvature associated with the conformal metric exhibits a periodic and highly structured spatial profile along the grain boundary. Notably, there are well-defined lines at \(x = (n + 1/2)a\) where the curvature inverts sign, producing alternating regions of positive and negative curvature aligned along the defect line. These sign changes arise from the structure of the numerator in the linear term and reflect the spatial arrangement of disclination dipoles. The curvature oscillates along the \(x\)-direction with spatial frequency \(2\pi/a\), matching the periodicity of the underlying defect distribution. These oscillations are most pronounced near \(y = 0\), where the curvature reaches its maximal amplitude. In addition, nonlinear contributions proportional to \(m^2\) produce sharply localized peaks centered on each disclination, enhancing both the intensity and confinement of the curvature near the core. This behavior is consistent with the enhanced strain fields observed in atomistic simulations of tightly packed 5|7 defect cores~\cite{Lorentzen2024}. The resulting curvature pattern resembles a comb-like structure of alternating peaks and valleys that faithfully reproduces the geometric features of periodic grain boundaries, as reported in simulations of graphene~\cite{Malola2010}.
\subsection{Asymptotic Regime of the Curvature}

Far from the grain boundary, the conformal factor decays linearly with \(|y|\), and the curvature exhibits exponential suppression. In the asymptotic regime \(|y| \gg a\), the scalar curvature~\ref{RicciR} behaves as

\begin{equation}
R(x,y) \sim \frac{4\pi^2 m(1+m)}{a^2} \, e^{-4\pi |y|/a} \cos\left( \frac{2\pi x}{a} \right),
\label{ricciAP}
\end{equation}
indicating that the geometric deformation is tightly confined to a narrow strip around the defect line. Despite this localized curvature, the configuration remains globally neutral in a topological sense. The positive and negative contributions from each disclination dipole cancel out over a full period, such that the integrated curvature vanishes.

This ensures the absence of global curvature defects or conical singularities, in accordance with the flat Euclidean geometry of the background space. Accordingly, the grain boundary can be interpreted as a localized geometric perturbation with no net topological charge, consistent with the geometric defect formalism introduced by Katanaev and Volovich~\cite{Katanaev1992}.

\section{Holonomy and Deficit Angles}

Holonomy transformations provide an efficient framework for describing key geometrical features of two-dimensional systems. They capture the net effect of parallel transport along closed loops in curved spaces. When curvature is present, a vector transported around such a loop may return rotated relative to its initial orientation. This residual rotation encodes global geometric information and is directly related to the total curvature enclosed by the path~\cite{Carvalho2013}. In the context of defect geometry, holonomy offers a powerful means of quantifying the angular mismatch induced by topological defects, such as disclinations and grain boundaries. This mismatch is measured by a deficit angle, which reflects deviations from flat-space parallel transport and is intrinsically linked to the integrated curvature over the enclosed region.

To obtain the total deficit angle \(\chi\) induced by the geometry, we consider the parallel transport of a vector along a circular path of fixed radius \(R\), centered at the origin. In conformal geometry, the deficit angle is given by~\cite{Carvalho2013}:
\begin{equation}
\chi = - \int_0^{2\pi} R \left. \frac{\partial \Omega}{\partial r} \right|_{r=R} d\theta.
\end{equation}

In the present model, we adopt the asymptotic form of the conformal factor, which captures the leading-order contribution of a periodic array of disclination dipoles aligned along the $x$-axis and decaying in the $y$-direction. Rewriting in polar coordinates $(x, y) = (r\cos\theta, r\sin\theta)$, we obtain
\begin{equation}
\Omega(r, \theta) \sim -\frac{m}{2a} |r \sin\theta|,
\end{equation}
Which leads to the following expression  for the deficit angle:
\begin{equation}
\chi = \frac{2m}{a}
\end{equation}
This result offers a geometric interpretation of the misorientation angle between crystalline domains. In particular, the value of \(\chi\) closely matches the rotation angles reported in STM experiments on low-angle grain boundaries (LAGBs) in graphene~\cite{YazyevLouie2010}, enabling a direct identification between the conformal geometric model and atomistic configurations.

This result is consistent with the interpretation of \(\chi\) as the geometric phase acquired by a vector under parallel transport around a loop enclosing a distribution of disclinations. As emphasized by Yazyev and Louie~\cite{YazyevLouie2010}, even in the absence of global curvature, the geometry retains a nontrivial holonomy: a vector transported around a closed path enclosing disclination dipoles experiences a net rotation. This holonomy captures the rotational mismatch between crystalline domains and aligns with the misorientation observed in STM imaging and DFT simulations of graphene grain boundaries.

Accordingly, the expression \(\chi = 2m/a\) provides a quantitative and physically meaningful measure of the angular mismatch induced by a periodic array of topological defects. It encodes the integrated curvature within the loop and offers a geometric interpretation of misorientation that is consistent with both theoretical models and experimental findings.


Our approach bears structural similarities to previous models involving periodic arrays of line defects governed by logarithmic potentials. Notably, Letelier’s spacetime construction~\cite{Letelier2001,Berche2020},  both employ logarithmic solutions of the Poisson equation to generate delta-like curvature sources. While Letelier’s approach is embedded in the Einstein–Cartan theory and includes torsion via off-diagonal metric components—interpreting defects in a fully spacetime context—Berche et al.~focus on closed-form expressions for interaction energies in defect lattices using Jacobi theta functions. In contrast, our model is purely Riemannian and conformal, without explicit torsion or dislocations. The geometry arises from a prescribed distribution of disclinations encoded in a conformal factor, leading to an effective metric with analytically tractable curvature. Remarkably, features analogous to dislocations and spin emerge naturally from the angular arrangement of alternating disclinations. This enables a minimal yet robust geometric description of grain boundaries in two-dimensional crystals, such as graphene, and provides a complementary perspective to both atomistic simulations and energy-based approaches.

\begin{figure}[htbp]
    \centering
    \includegraphics[width=0.75\textwidth]{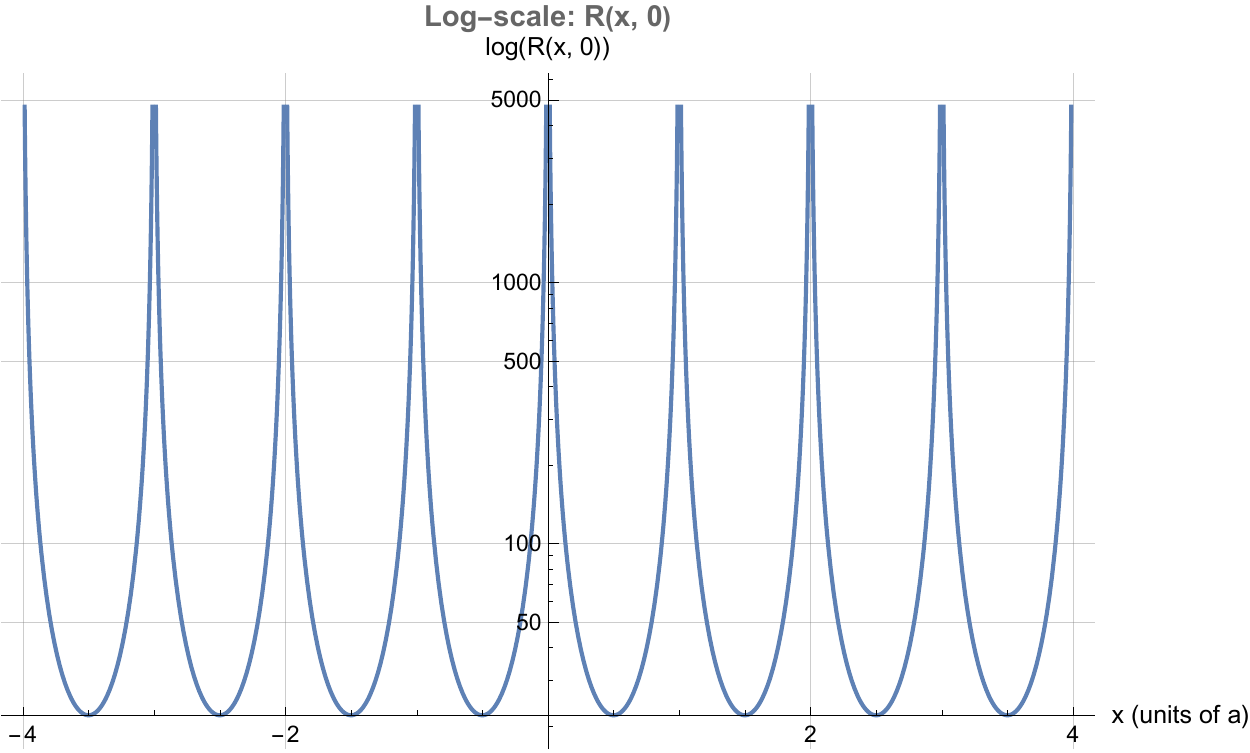}
    \caption{Logarithmic plot of the transverse curvature profile \( R(x, 0) \) for a periodic line of alternating disclinations. The use of log-scale reveals the periodic structure of sharp curvature peaks that would otherwise be numerically invisible due to their divergence at defect cores.}
    \label{fig:logplot_Rx0}
\end{figure}
Figure~\ref{fig:logplot_Rx0} presents the transverse curvature profile \( R(x, 0) \) using a logarithmic vertical scale. This log-scale representation is essential to resolve the multiple divergent peaks associated with the disclination cores located at \( x = na \), where \( n \in \mathbb{Z} \). In the original linear plot, these peaks diverge and dominate the range, masking the intermediate structure. The logarithmic scaling compresses the vertical axis, making the periodic nature of the curvature distribution manifest.

Physically, this confirms the expected periodic alternation of concentrated curvature around defect positions in the grain boundary, and highlights the singular nature of the geometric response in the conformal metric. The logarithmic plot thus provides a clear visualization of the curvature landscape, especially in regimes where standard linear plots fail to resolve spatial features with high gradients.
\begin{figure}[htbp]
    \centering
    \includegraphics[width=0.7\textwidth]{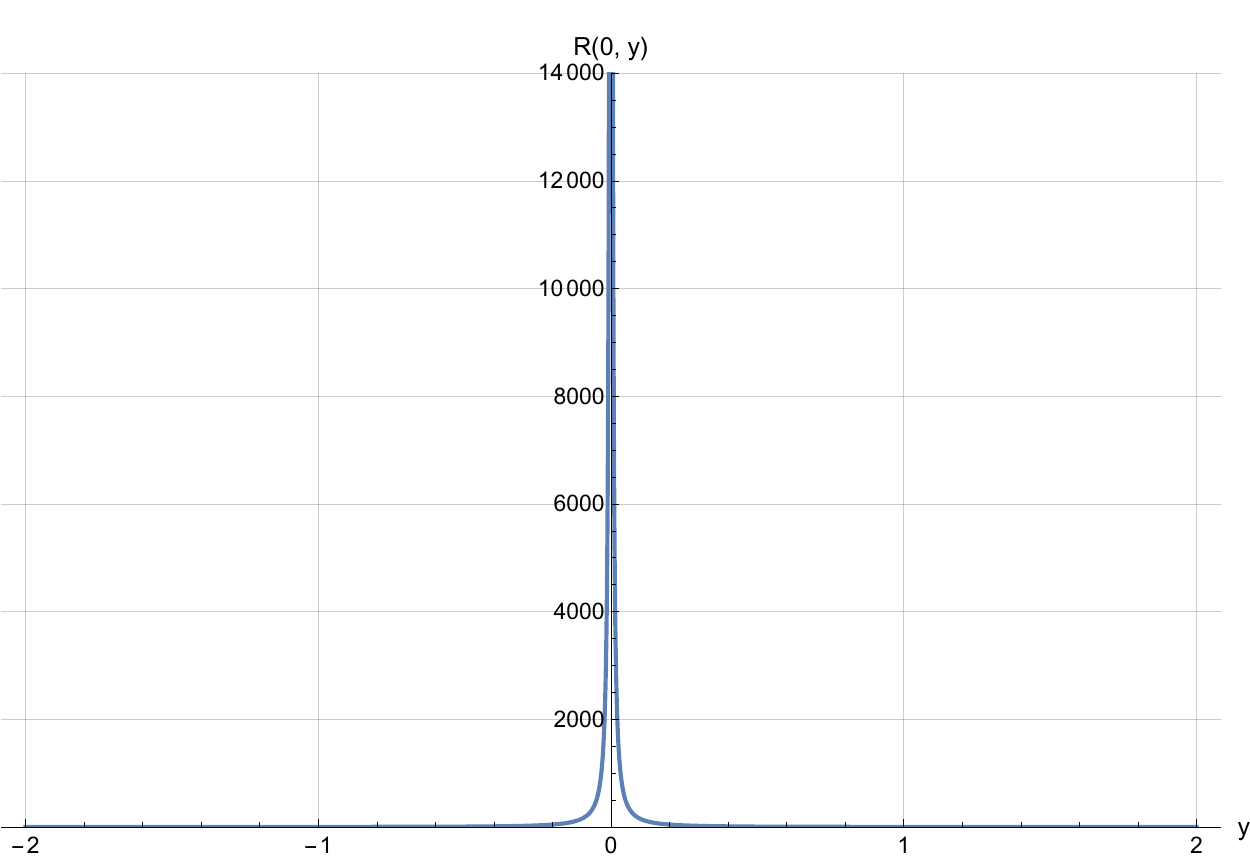}
    \caption{Linear plot of the scalar curvature \( R(0, y) \) as a function of the transverse coordinate \( y \). The curvature exhibits a sharp peak at the defect line (\( y = 0 \)), reflecting strong geometric localization.}
    \label{fig:R0y_linear}
\end{figure}

\begin{figure}[htbp]
    \centering
    \includegraphics[width=0.7\textwidth]{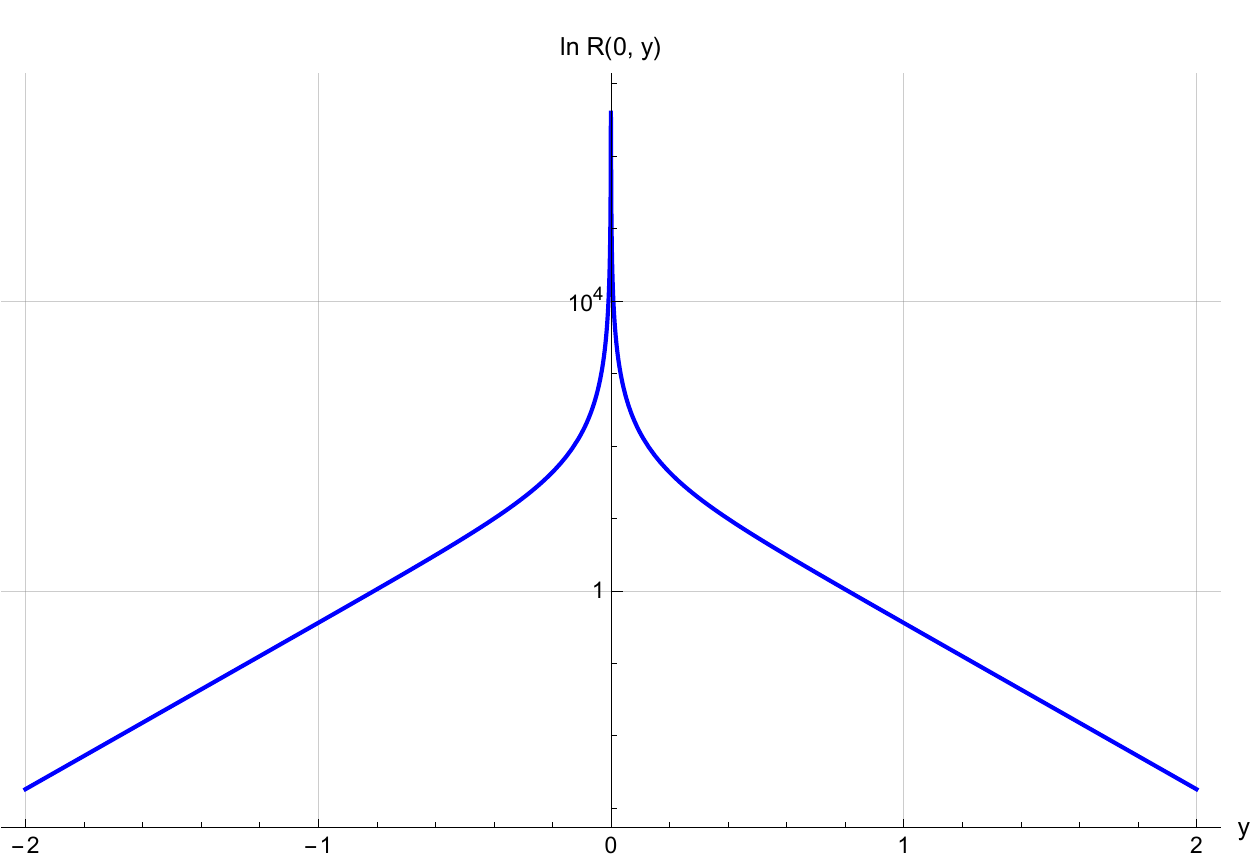}
    \caption{Log-log plot of the scalar curvature \( \ln R(0, y) \). The nearly linear decay on both sides of the central peak confirms the exponential suppression of curvature away from the grain boundary. This behavior indicates geometric confinement.}
    \label{fig:R0y_log}
\end{figure}
Figures~\ref{fig:R0y_linear} and~\ref{fig:R0y_log} complement each other in describing the transverse behavior of the scalar curvature \( R(0, y) \) near a periodic array of disclinations.

In Figure~\ref{fig:R0y_linear}, the linear plot shows a sharp and narrow peak centered at \( y = 0 \), indicating intense geometric deformation localized near the grain boundary. However, due to the high dynamic range of the curvature field, details of the decay away from the core are visually suppressed.

Figure~\ref{fig:R0y_log} uses a log-log scale, which reveals a quasi-linear decay profile. This confirms that the curvature field decays approximately exponentially in \( |y| \), consistent with geometric confinement induced by a conformal deformation. The symmetry of the decay on both sides also reflects the underlying parity of the defect distribution.

Together, these visualizations establish that the conformal model accurately captures both the localization and decay structure of the curvature field around a periodic disclination array, supporting its use as an effective geometric model for grain boundary systems.

There exists a deep analogy between the Gauss--Bonnet theorem and the holonomy angle acquired by a vector after parallel transport around a closed loop. This relation, explored by Vickers~\cite{Vickers1987}, expresses the holonomy angle \( \chi \) as a flux of curvature through the enclosed region:
\begin{equation}
\chi = \int_{\partial Q} k_g \, ds = \int_Q K \, dA,
\end{equation}
where \( k_g \) is the geodesic curvature along the boundary \( \partial Q \), and \( K \) is the Gaussian curvature over the region \( Q \). 

This geometric phase admits a direct analogue in the Aharonov--Bohm effect, where the phase shift experienced by a charged particle is given by
\begin{equation}
\chi_{\text{AB}} = \frac{ie}{\hbar c} \oint \vec{A} \cdot d\vec{l} = \frac{ie}{\hbar c} \Phi,
\end{equation}
with \( \vec{A} \) being the magnetic vector potential and \( \Phi \) the magnetic flux through the surface enclosed by the loop. In this analogy, the curvature \( K \) plays the role of a geometric flux, highlighting the topological nature of holonomy in curved spaces.

\section{Comparison with Atomistic Models}

The conformal geometric model developed here provides an analytically tractable continuum description of grain boundaries in graphene, represented as periodic arrays of disclination dipoles. To explore its physical relevance, we qualitatively compare its predictions with atomistic simulations, particularly those reported in Refs.~\cite{YazyevLouie2010,Malola2010,Lorentzen2024}. Core features such as structural periodicity, domain misorientation, and localized deformation appear consistently across both approaches.

In their seminal study, Yazyev and Louie~\cite{YazyevLouie2010} classified low-angle grain boundaries (LAGBs) in graphene as periodic sequences of 5|7 disclination dipoles forming (1,0) and (0,1) dislocations. Two energetically favorable configurations were identified: LAGB I, with a linear chain of (1,0) dislocations aligned along the armchair direction (\(\theta = 21.8^\circ\)), and LAGB II, a denser arrangement involving both dislocation types (\(\theta = 32.2^\circ\)). In our model, these structures are encoded by periodic dipoles of strength \(\pm m\), spaced by a distance \(a\) along the \(x\)-axis.

By identifying \(m = \pi/3\), consistent with isolated pentagonal or heptagonal defects, the holonomy angle \(\chi = 2m/a\) can be compared directly with the experimentally observed misorientation angles, assuming the qualitative correspondence \(\chi \approx \theta\). This leads to the following estimates for the defect spacing:
\begin{itemize}
    \item For LAGB I (\(\theta \approx 0.3805\) rad), \(a \approx 5.5\);
    \item For LAGB II (\(\theta \approx 0.5618\) rad), \(a \approx 3.73\).
\end{itemize}

These values are consistent with the order of magnitude observed in atomistic simulations, where (5|7) dipoles are separated by multiples of graphene’s lattice constant.

In the atomistic framework, the misorientation angle is governed by the spacing \(d\) between dislocations and follows Frank’s formula, \(\theta = 2 \arcsin\left( |\vec{b}| / 2d \right)\). Within the conformal model, the angle arises naturally as the holonomy associated with parallel transport around the defect array~\cite{Carvalho2013}. The spacing \(a\) corresponds to \(d\), establishing a qualitative mapping between the discrete and continuous parameters.

Beyond misorientation, the model also captures the spatial localization of geometric deformation. Although we do not compute elastic energies directly, the conformal factor \(\Omega(x,y)\) plays the role of a geometric potential, analogous to the gravitational potential in $(2+1)$-dimensional gravity~\cite{Katanaev1992}. It governs local metric dilation and acts as a scalar measure of strain.

This interpretation is further supported by the continuum elastic theory of defects developed by Berche et al.~\cite{Berche2020}, where scalar potentials—analogous to our conformal factor—encode the local distortion induced by disclinations. These potentials determine the structure of the induced metric through conformal transformations, and the corresponding elastic energy densities, although derived from the gradients of the potential, exhibit the same spatial localization and periodic modulation as \(\Omega(x,y)\).

In this context, the deformation energy density may be qualitatively expressed as~\cite{Carvalho2004,Andrade2004PRD,Berche2020}
\begin{equation}
\mathcal{E}(x,y) \propto \Omega(x,y),
\end{equation}
which serves as a descriptor of the geometric intensity of deformation. While this expression does not define a quantitative energy functional, it captures the spatial regions of high and low strain. This qualitative behavior mirrors the peaks and troughs observed in first-principles simulations~\cite{YazyevLouie2010,Lorentzen2024}, reinforcing the role of \(\Omega(x,y)\) as an effective scalar field summarizing both the geometric and physical structure of grain boundaries in two-dimensional materials.

In summary, the conformal model captures key geometric features of grain boundaries in graphene—such as misorientation, periodic structure, and localized strain—through a continuous and analytically solvable framework. Its qualitative agreement with atomistic simulations supports the usefulness of conformal geometry for describing strain fields and informing continuum models of two-dimensional materials.

\begin{figure}[h!]
    \centering
    \includegraphics[width=0.6\textwidth]{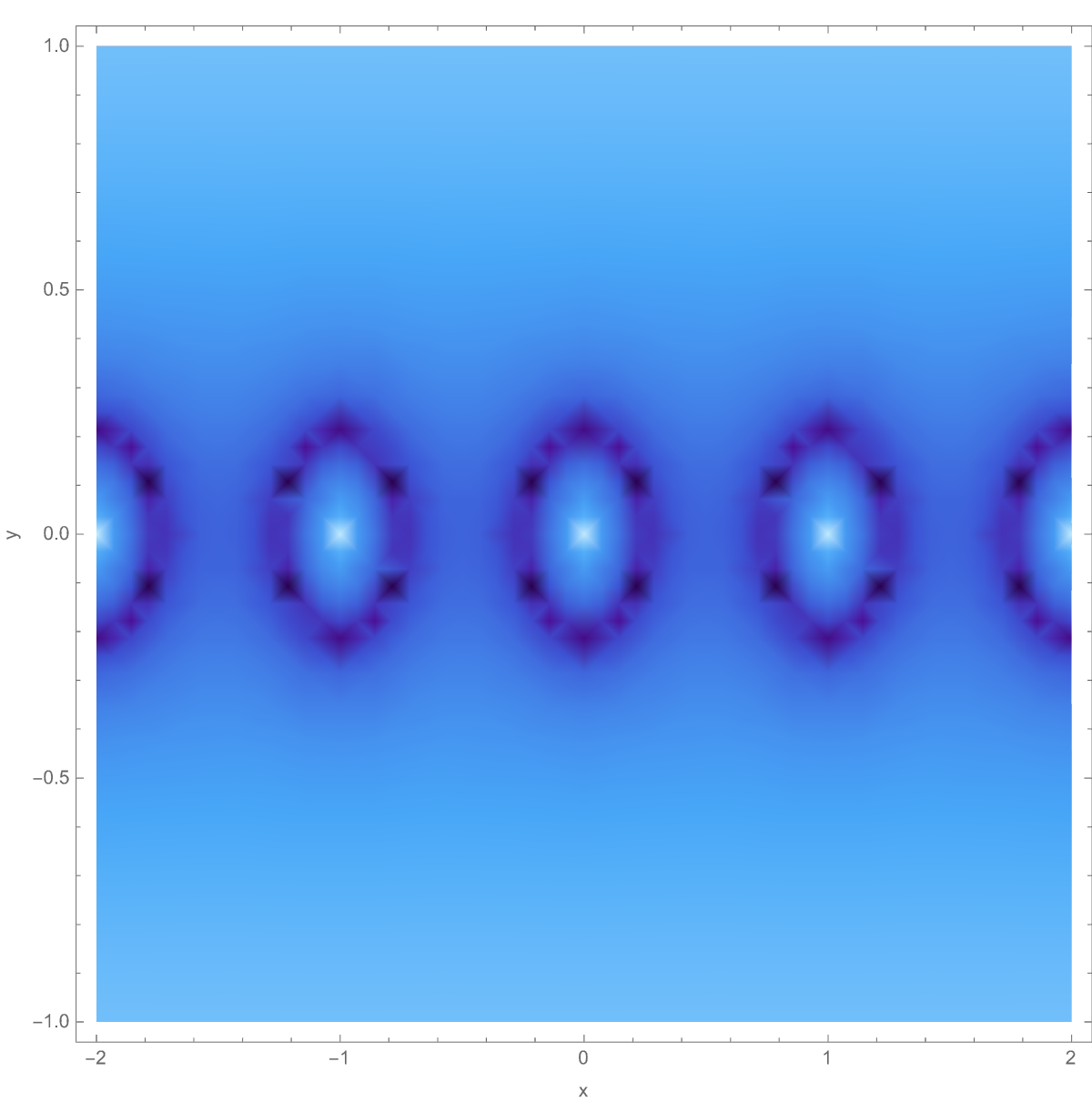}
    \caption{Spatial distribution of the deformation energy density $\mathcal{E}(x,y) \propto \Omega(x,y)$ for a periodic array of disclination dipoles along the $x$-axis. The figure displays localized peaks in energy near the defect cores at $y = 0$ and shows rapid decay away from the grain boundary. This pattern highlights the spatial confinement and periodic modulation of geometric deformation in the conformal model.}
    \label{fig:energy_density_map}
\end{figure}
The energy density map in Figure~\ref{fig:energy_density_map} illustrates two key features of the geometric deformation encoded by the conformal factor $\Omega(x,y)$. First, the energy is strongly confined near the defect line at $y = 0$, exhibiting a rapid exponential decay in the transverse ($y$) direction. This reflects the localized nature of the curvature induced by the periodic disclination dipoles. Second, the distribution shows a periodic modulation along the $x$-axis, with peaks regularly spaced according to the inter-defect separation. This spatial structure mirrors the geometry imposed by the grain boundary configuration and highlights the ability of the conformal model to capture the anisotropic and highly localized character of deformation in two-dimensional crystals. Together, these features validate the use of $\Omega(x,y)$ as a continuous geometric field capable of reproducing qualitative aspects of deformation energy observed in atomistic simulations~\cite{YazyevLouie2010,Lorentzen2024}.
\begin{figure}[ht]
    \centering
    \includegraphics[width=0.65\textwidth]{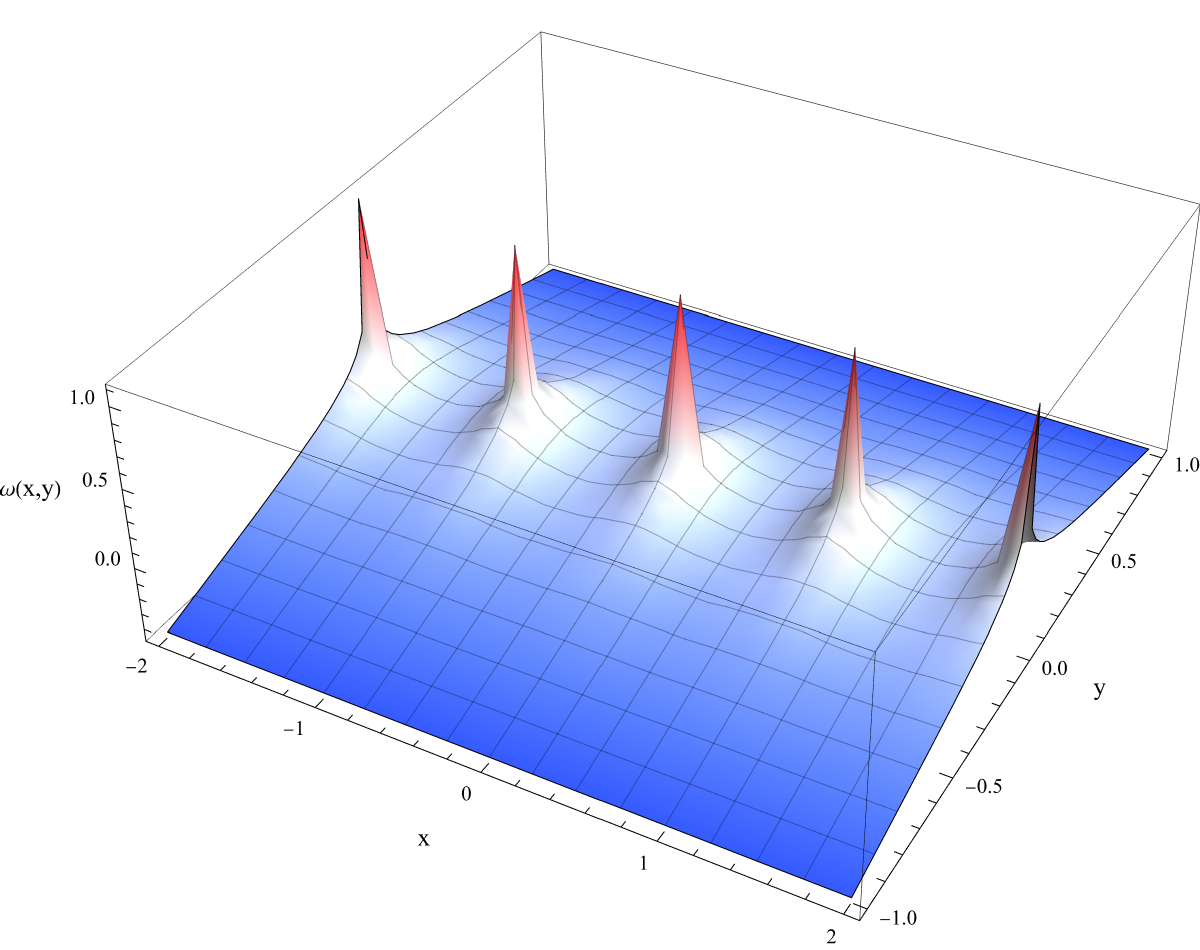}
  \caption{Energy surface derived from the conformal factor $\Omega(x,y)$. Although the underlying function is the same as in Figure~\ref{fig:omega_3d_grain}, here it is interpreted as a proxy for the deformation energy density, emphasizing its physical role in capturing the spatial localization and modulation of strain.}

    \label{fig:energy_surface}
\end{figure}
The energy surface shown in Figure~\ref{fig:energy_surface} illustrates the spatial structure of the conformal factor $\Omega(x,y)$, adopted here as a proxy for the deformation energy density. The figure reveals localized peaks aligned along the grain boundary ($y = 0$), reflecting the periodic arrangement of disclination dipoles. The rapid decay in the transverse direction demonstrates geometric confinement of the deformation field. This pattern closely mirrors the energy modulation observed in atomistic simulations of grain boundaries and reinforces the interpretation of $\Omega(x,y)$ as a physically meaningful scalar field capturing both curvature and localized strain in a unified geometric approach. While our model focuses on a continuous geometric field, Berche et al.~\cite{Berche2020} evaluate defect interactions via regularized lattice sums expressed through Jacobi theta functions. While their approach relies on discrete summation over defect interactions, ours encodes geometric effects continuously via a conformal field, enabling analytical access to curvature and holonomy.

Beyond the classical continuum description, the geometric framework developed here offers promising extensions to quantum systems. The metric induced by the conformal field provides a natural background to investigate Landau level quantization in the presence of topological defects~\cite{Garcia2017}. The spatial localization encoded in the deformation pattern may also model effective confinement potentials for quantum dots or wells in two-dimensional materials~\cite{Bueno2014}. Furthermore, this class of metrics can serve as curved backgrounds for quantum fields, enabling the exploration of curvature-induced phases and geometric couplings in strained media~\cite{Bueno2012}. These perspectives suggest a fertile interface between classical geometry and quantum phenomena in low-dimensional systems.

\section{Conclusions}

We presented a geometric model for grain boundaries in graphene based on a periodic array of disclination dipoles, formulated within the framework of two-dimensional conformal geometry. By solving a reduced form of the Einstein equations with alternating curvature sources, we obtained an explicit expression for the conformal factor \( \Omega(x, y) \), from which the induced metric, scalar curvature, and holonomy were derived analytically.

The resulting geometry captures key features observed in atomistic simulations, including localized curvature along the defect line, periodic strain modulation, and angular misorientation. In particular, the scalar curvature exhibits a comb-like profile, with exponential decay transverse to the defect and periodicity consistent with the 5|7 structure. The integrated curvature vanishes over a period, ensuring topological neutrality, while the holonomy encodes the misorientation angle between adjacent crystalline domains.

A central result of this work is the analytic construction of a continuous metric deformation that models realistic grain boundary configurations such as LAGB I and II~\cite{YazyevLouie2010}. This provides a direct link between discrete atomic-scale defects and a continuum geometric description, in contrast to purely numerical or torsion-based approaches~\cite{Letelier2001}. Our model relies solely on Riemannian geometry and depends on two tunable parameters: the dipole strength \( m \) and the spacing \( a \), enabling controlled exploration of geometric effects.

These results demonstrate that the conformal metric model captures key features reported in atomistic studies of graphene, reinforcing its utility as a bridge between continuum geometric theory and discrete material simulations.

We also proposed interpreting the curvature field as a proxy for elastic energy density, observing qualitative agreement with energy profiles obtained from density functional theory~\cite{Lorentzen2024,Malola2010}. This geometric interpretation opens new avenues for understanding the mechanical and electronic consequences of defect patterns, with potential applications in strain engineering and topological phenomena in graphene and related materials.

Overall, this work provides a coherent and analytically tractable framework for modeling extended topological defects in two-dimensional crystals. Future developments may include the incorporation of torsion to describe dislocations explicitly, the coupling of electronic degrees of freedom to curved geometries, and the extension to more complex or disordered defect networks. Geometric approaches of this kind can also complement data-driven structure prediction methods recently applied to other two-dimensional systems such as silicene~\cite{Zhang2023Silicene}.

{\bf Acknowledgments:} C. Furtado has been supported by the CNPq (project PQ Grant 1A No. 311781/2021-7).

\bibliographystyle{iopart-num}
  \bibliography{biblio} 
\end{document}